\documentclass[runningheads]{llncs}

\usepackage{graphicx,amsmath,xcolor,cite,wrapfig,bbm,comment, subfig}
\usepackage{array,multicol,booktabs,tabularx,todonotes,dsfont,multirow,caption,hyphenat}
\usepackage[inline]{enumitem}

\usepackage[breaklinks=true,bookmarks=false]{hyperref}
\hypersetup{
	colorlinks=true,
	linkcolor=blue,
	urlcolor=magenta,
}

\usepackage{floatrow}
\newfloatcommand{capbtabbox}{table}[][\FBwidth]

\newcolumntype{P}[1]{>{\centering\arraybackslash}p{#1}}

\def\etal{{\textit{et al. }}}
\def\covid{{COVID-19 }}
\def\xray{{X-Ray }}

\begin{document}
\title{CovidAID: COVID-19 Detection Using Chest X-Ray}
\titlerunning{CovidAID : COVID-19 AI Detector}
\author{Arpan Mangal$^1$ \and Surya Kalia$^1$ \and Harish Rajgopal$^2$ \and Krithika Rangarajan$^{3,1}$ \and Vinay Namboodiri$^{2,4}$ \and Subhashis Banerjee$^1$ \and Chetan Arora$^1$}
\authorrunning{Mangal et al.}
\institute{
	Indian Institute of Technology Delhi \and 
	Indian Institute of Technology Kanpur \and 
	All India Institute of Medical Sciences Delhi \and
	University of Bath, UK
}

\maketitle    

\begin{center}
\url{https://github.com/arpanmangal/CovidAID}
\end{center}

\begin{abstract}
The exponential increase in \covid patients is overwhelming healthcare systems across the world. With limited testing kits, it is impossible for every patient with respiratory illness to be tested using conventional techniques (RT-PCR). The tests also have long turn-around time, and limited sensitivity. Detecting possible COVID-19 infections on Chest \xray may help quarantine high risk patients while test results are awaited. \xray machines are already available in most healthcare systems, and with most modern \xray systems already digitized, there is no transportation time involved for the samples either. In this work we propose the use of chest \xray to prioritize the selection of patients for further RT-PCR testing. This may be useful in an inpatient setting where the present systems are struggling to decide whether to keep the patient in the ward along with other patients or isolate them in \covid areas. It would also help in identifying patients with high likelihood of COVID with a false negative RT-PCR who would need repeat testing.  Further, we propose the use of modern AI techniques to detect the \covid patients using X-Ray images in an automated manner, particularly in settings where radiologists are not available, and help make the proposed testing technology scalable. We present CovidAID: \covid AI Detector, a novel deep neural network based model to triage patients for appropriate testing. On the publicly available  covid-chestxray-dataset~\cite{cohen2020covid} dataset, our model gives $90.5\%$ accuracy with $100\%$ sensitivity (recall) for the \covid infection. We significantly improve upon the results of Covid-Net~\cite{wang2020covidnet} on the same dataset.

\keywords{Deep learning, \covid Detection, Chest \xray}
\end{abstract}

\section{Introduction}

The sudden spike in the number of patients with \covid, a new respiratory virus, has put unprecedented load over healthcare systems across the world. In many countries, the healthcare systems have already been overwhelmed. There are limited kits for diagnosis, limited hospital beds for admission of such patients, limited personal protective equipment (PPE) for healthcare personnel and limited ventilators. It is thus important to differentiate which patients with severe acute respiratory illness (SARI) could have \covid infection in order to efficiently utilize the limited resources. In this work we propose the use of chest X-Ray to detect \covid infection in the patients exhibiting symptoms of SARI. Using our tool one can classify a given \xray in one of the four classes: normal, bacterial pneumonia, viral pneumonia, and covid pneumonia. The use of \xray has several advantages over conventional diagnostic tests:
\begin{enumerate}[leftmargin=*]
	\item \xray imaging is much more widespread and cost effective than the conventional diagnostic tests.
	\item Transfer of digital \xray images does not require any transportation from point of acquisition to the point of analysis, thus making the diagnostic process extremely quick. 
	\item Unlike CT Scans, portable \xray machines also enable testing within an isolation ward itself, hence reducing the requirement of additional Personal Protective Equipment (PPE), an extremely scarce and valuable resource in this scenario. It also reduces the risk of hospital acquired infection for the patients.
\end{enumerate}

The main contribution of this work is in proposing a novel deep neural network based model for highly accurate detection of \covid infection from the chest \xray images of the patients. Radiographs in the current setting are in most cases interpreted by non-radiologists. Further, given the novelty of the virus, many of the radiologists themselves may not be familiar with all the nuances of the infection, and may be lacking in the adequate expertise to make highly accurate diagnosis. Therefore this automated tool can serve as a guide for those in the forefront of this analysis.

We would like to re-emphasize that we are not proposing the use of the proposed model as  alternative to the conventional diagnostic tests for \covid infection, but as a triage tool to determine the suitability of a patient with SARI to undergo the test for \covid infection.

To help accelerate the research in this area, we are releasing our training code and trained models publicly for open access at \url{https://github.com/arpanmangal/CovidAID}. However, we note that both the model and this report merely captures our current understanding of this rapidly evolving problem, that too on very limited data currently available. We will keep updating the model and this report as we get newer understanding and better results.

\section{Related Work}

\subsection{Pneumonia detection in Chest X-Rays}
Various deep learning based approaches have been developed to identify different thoracic diseases, including pneumonia ~\cite{rajpurkar2017chexnet, yaonet, wang2017chestx,ranjanetal}. We choose CheXNet~\cite{rajpurkar2017chexnet} to build upon, which could detect pneumonia from chest X-Rays at a level exceeding practicing radiologists. CheXNet is trained on ChestX-ray14~\cite{wang2017chestx} (the largest publicly available chest X-ray dataset), gives better performance than previous approaches~\cite{yaonet, wang2017chestx}, and has a  simpler architecture than later approaches~\cite{ranjanetal}.

CheXNet~\cite{rajpurkar2017chexnet} is a 121-layer DenseNet \cite{huang2017densely} based model trained on the ChestX-ray14~\cite{wang2017chestx} dataset comprising of 112,120 frontal-view chest \xray images. The model is trained to classify \xray images into 14 different thoracic disease classes, including pneumonia. Given the visual similarity of the input samples, we found this to be the closest pre-trained backbone to develop a model for identifying \covid pneumonia.

\subsection{\covid detection in Chest X-Rays}
 Since the recent sudden surge of COVID-19 infections across the world, many alternative screening approaches have been developed to identify suspected cases of COVID-19. However there are only  limited such open-source applications  available for use~\cite{smartct, wang2020covidnet, covid19_VGG} which use chest X-Ray images.   Publicly available data on chest X-Rays for COVID-19 are also limited.
 
 
 COVID-Net~\cite{wang2020covidnet} is the only approach having an open source and actively-maintained tool which has ability to identify COVID-19 as well as other pneumonia while showing considerable sensitivity for COVID-19 detection. COVID-Net uses a machine-driven design exploration to learn the architecture design starting from initial design prototype and requirements. It takes as input a chest X-Ray image and outputs a prediction among three classes: \textit{Normal}, \textit{Pneumonia} and \textit{COVID-19}. We treat this model as our baseline, comparing our results with it.
 

\section{Method}

Given the limited amount of available \xray samples (including bacterial and viral pneumonia), it is hard to train a deep neural network from scratch. Hence we use a pre-trained backbone trained on a large dataset. Our approaches use the pre-trained model of CheXNet~\cite{rajpurkar2017chexnet}, released by Weng \etal \cite{chexnet_implementation}.


\subsection{Problem Formulation and Loss Function}

We aim to classify a given frontal-view chest \xray image into the following classes: \textit{Normal}, \textit{Bacterial Pneumonia}, \textit{Viral Pneumonia} and \textit{COVID-19}. We have trained our model in two configurations, the one which classifies into the above four classes, and the other configuration with three classes (clubbing viral and bacterial pneumonia into one). The motivation behind the four class configuration is to better understand if any confusion between regular pneumonia and \covid is due to the similarity of pathology between \covid and viral pneumonia.


\begin{figure}[!ht]
	\centering
	\captionsetup{justification=centering}
	\subfloat[Normal Lung]{\includegraphics[width=0.23\linewidth, height=0.23\linewidth]{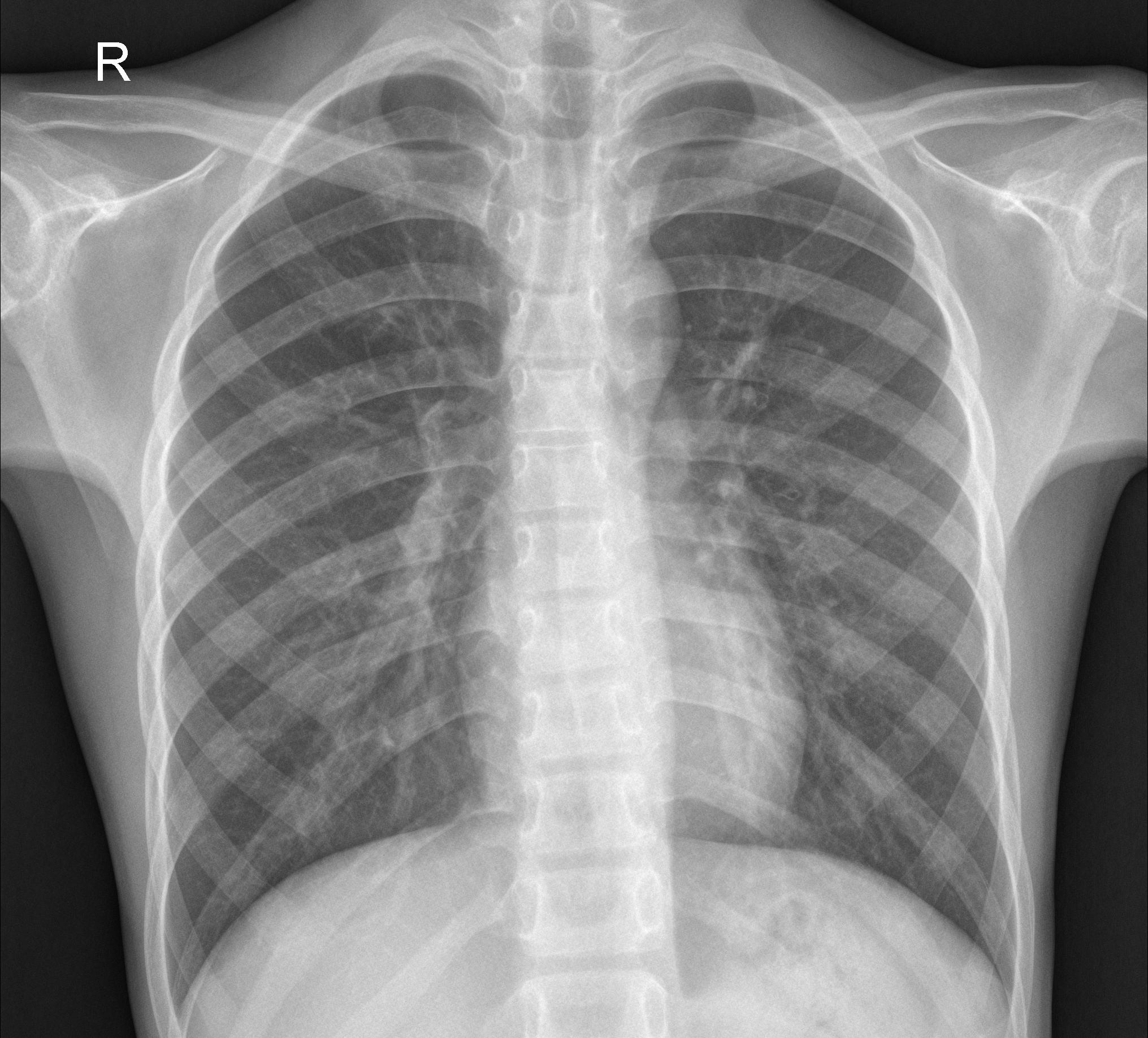}}
	\hfill
	\subfloat[Bacterial  Pneumonia]{\includegraphics[width=0.23\linewidth, height=0.23\linewidth]{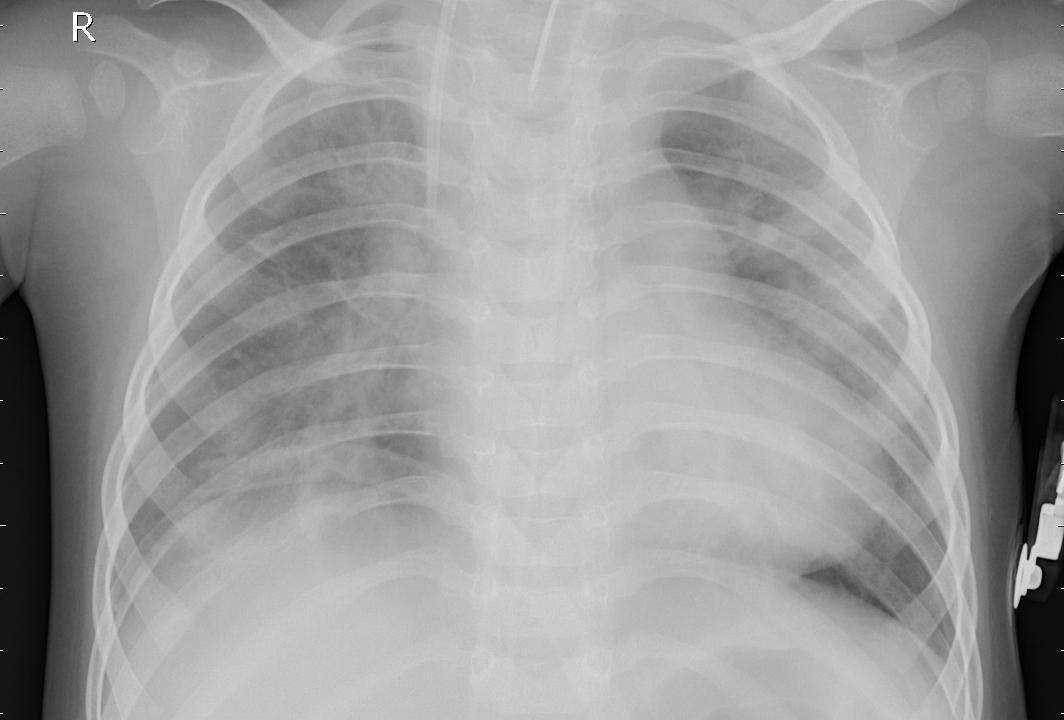}}
	\hfill
	\subfloat[Viral Pneumonia]{\includegraphics[width=0.23\linewidth, height=0.23\linewidth]{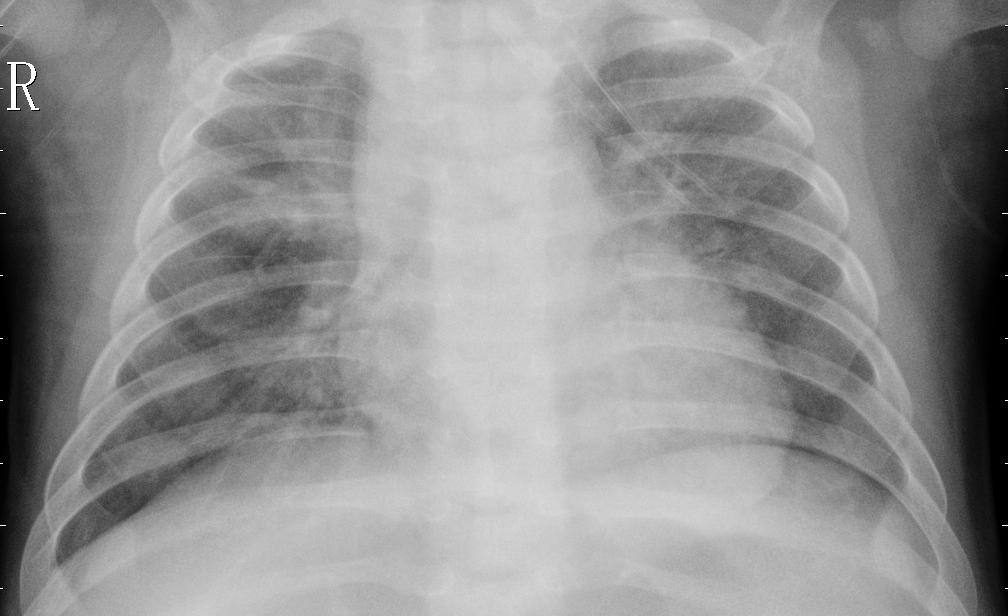}}
	\hfill
	\subfloat[\covid]{\includegraphics[width=0.23\linewidth, height=0.23\linewidth]{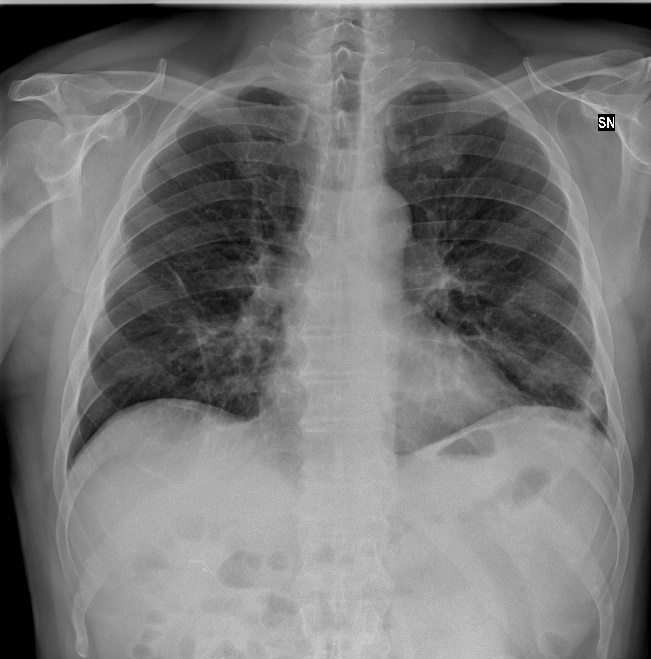}}

	\caption{Class-wise examples of frontal-view chest X-Ray images. The patient with bacterial pneumonia shows consolidation of the right lower lobe. Radiograph of a patient with viral pneumonia shows patchy consolidation in the right middle and lower zone. The last image from a patient with COVID pneumonia shows patchy ground glass opacity in the Left Lower Zone}
	\label{fig:ex_imgs}
\end{figure}

Similar to CheXNet~\cite{rajpurkar2017chexnet}, we treat each class as a binary classification problem, with input as a frontal-view chest \xray image $X$ and output being a binary labels $y_c \in \{ 0, 1 \}$, indicating absence or presence of class $c$ symptoms in the image respectively. We use the weighted binary cross-entropy loss as suggested by CheXNet~\cite{rajpurkar2017chexnet}:
\begin{align}
\mathcal{L}(X, y; \theta) = \sum_{c=1}^{C} 
& \left( - w_c^+ \mathbbm{1} \{ y = c \} \log p_c(\hat{y}=1 \mid X;\theta) \right. \nonumber \\
& ~ \left. - w_c^- \mathbbm{1} \{ y \neq c \} \log p_c(\hat{y}=0 \mid X;\theta) \right).
\end{align}
Here, $C$ is the number of classes and $y$ is the ground truth for $X$. $p_c(\hat{y}=1 \mid X;\theta)$ and $p_c(\hat{y}=0 \mid X;\theta)$ are the probability scores for $X$ being and not being in class $c$, respectively, as assigned by the network based on network weights $\theta$. The two terms are weighted by $w_c^+ = \frac{N_c}{N_c+P_c}$ and $w_c^- = \frac{P_c}{N_c+P_c}$, where $P_c$ and $N_c$ are the number of positive and negative samples of class $c$, respectively in the training set.

\subsection{Model Architecture}

Our model contains pre-trained CheXNet~\cite{rajpurkar2017chexnet}, with a 121-layer Dense Convolutional Network (DenseNet)~\cite{huang2017densely} backbone, followed by a fully connected layer. We replace CheXNet's~\cite{rajpurkar2017chexnet} final classifier of 14 classes with our classification layer of 4 classes (3 classes for the clubbed pneumonia configuration), each with a sigmoid activation to produce the final output.

\begin{figure}[!htb]
    \centering
    \includegraphics[width=\linewidth]{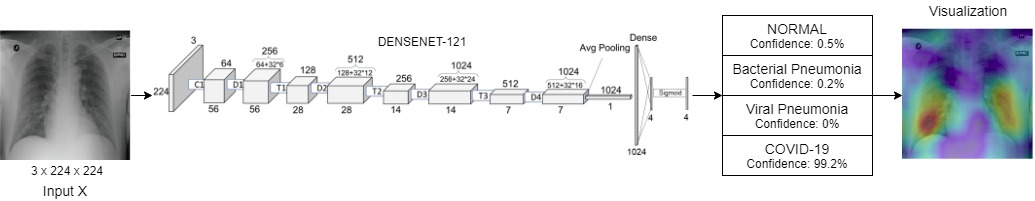}
    \caption{CovidAID model (DenseNet image from~\cite{densenetimg})}
    \label{fig:CovidAID}
\end{figure}

\subsection{Training}

For training we initialize our model with pre-trained weights from CheXNet implementation by Weng \etal \cite{chexnet_implementation}, and then following the two stage training process described below:
\begin{enumerate}[leftmargin=*]
\item In the first step, DenseNet's backbone weights are frozen and only the final fully connected layer is trained. Training is performed using Adam optimizer with following parameters: $\beta_1 = 0.9$, $\beta_2 = 0.999$, and learning rate $10^{-4}$. We use mini-batches of size 16, and train for about 30 epochs. The model with the lowest validation loss is selected for next stage.
\item In the second stage, the network weights are initialized from above, but the whole network is trained end-to-end (all layers), using the same hyper-parameters. We use mini-batch size of 8 in this stage due to memory constraints, and train for 10 epochs. Again, the model with lowest validation loss is selected for testing.
\end{enumerate}

\section{Dataset and Evaluation}

We use the covid-chestxray-dataset \cite{cohen2020covid} for \covid frontal-view chest \xray images and chest-xray-pneumonia dataset \cite{kpneumonia} for frontal-view chest \xray images with bacterial/viral pneumonia as well as of normal lungs. We use the pre-trained CheXNet model, thus implicitly using robust features obtained after training on ChestX-ray14~\cite{wang2017chestx} dataset.

The covid-chestxray-dataset ~\cite{cohen2020covid} does not contain proper data split for training purposes, so we perform our own split, as shown in Tables \ref{datasplit} and \ref{datasplit_patient}. Since multiple images for the same patient could be found in the dataset, we split the data by patient-IDs to prevent any information leakage. We choose $20\%$ of the images as test set, and 10 images are kept as validation set.  

\begin{table}
\setlength\tabcolsep{10pt}
\centering
\begin{tabular}{l c c c c c}
	\toprule[1.5pt]
	& \multirow{2}{*}{Normal} & \multicolumn{2}{c}{Pneumonia} & \multirow{2}{*}{\covid} & \multirow{2}{*}{Total} \\
	\cmidrule[0.5pt]{3-4}
	& & Bacterial & Viral & & \\
	\midrule[1pt]
	Train & 1341 & 2530 & 1337 & 115 & 5323 \\
	Val & 8 & 8 & 8 & 10 & 37 \\
	Test & 234 & 242 & 148 & 30 & 654 \\
	\bottomrule[1.5pt] 
\end{tabular}
\caption{Sample-wise Data Split}
\label{datasplit}
\end{table}

\begin{table}
	\setlength\tabcolsep{10pt}
	\centering
	\begin{tabular}{l c c c c c}
		\toprule[1.5pt]
		& \multirow{2}{*}{Normal} & \multicolumn{2}{c}{Pneumonia} & \multirow{2}{*}{\covid} & \multirow{2}{*}{Total} \\
		\cmidrule[0.5pt]{3-4}
		& & Bacterial & Viral & & \\
		\midrule[1pt]
		Train & 1000 & 1353 & 1083 & 80 & 3516 \\
		Val & 8 & 7 & 7 & 7 & 29 \\
		Test & 202 & 77 & 126 & 19 & 424 \\
		\bottomrule[1.5pt]
	\end{tabular}
	\caption{Patient-wise Data Split}
	\label{datasplit_patient}
\end{table}

\subsection{Sampling}

The combined dataset (covid-chestxray-dataset and chest-xray-pneumonia) has a high data imbalance due to scarce \covid data. Note that this imbalance is different from the positive-negative class imbalance (for which $w^+ \text{ and } w^-$ were introduced in the loss function). To ensure that training loss due to \covid does not get masked by training loss due to other classes, we consider only a random subset of pneumonia data in each batch. The size of this subset should neither be too small, which will lead to overfitting on the COVID-19 data, nor too large to mask the \covid loss, and is fixed empirically. In each batch we take data from classes \textit{Normal}, \textit{Bacterial Pneumonia}, \textit{Viral Pneumonia} and \textit{\covid} in the ratio $5:5:5:1$. In case of the three class classification network, this ratio is $7:7:1$.

\section{Results}

Our results indicate that this approach can lead to \covid detection from X-Ray images with an AUROC (Area under ROC curve) of 0.9994 for the \covid positive class, with a mean AUROC of 0.9738 (for 4-class classification configuration). Since we have modeled the problem as a binary classification problem for each class, given an input image $X$, we treat the class with maximum confidence score as the final prediction for calculating Accuracy, Sensitivity (Recall), PPV and confusion matrix.

\begin{figure}[ht]
\begin{floatrow}
\capbtabbox{%
\centering
\begin{tabular}{l c c c}
	\toprule[1.5pt]
	Pathology & AUROC & Sensitivity & PPV\\
	\midrule[1pt]
	Normal Lung & 0.9795 & 0.744 & 0.989 \\
	\multirow{2}{*}{Pneumonia} & \multirow{2}{*}{0.9814} & \multirow{2}{*}{0.995} & \multirow{2}{*}{0.868} \\
	& & & \\
	\covid & 0.9997 & 1.000 & 0.968  \\
	\bottomrule[1.5pt]
\end{tabular}
}{%
	\caption{Class-wise results for 3-class classification}
	\label{results3}
}
\capbtabbox{%
\centering
\begin{tabular}{l c c c}
	\toprule[1.5pt]
	Pathology & AUROC & Sensitivity & PPV\\
	\midrule[1pt]
	Normal Lung & 0.9788 & 0.761 & 0.989 \\
	Bact. Pneumonia & 0.9798 & 0.961 & 0.881 \\
	Viral Pneumonia & 0.9370 & 0.872 & 0.721 \\
	\covid & 0.9994 & 1.000 & 0.938  \\
	\bottomrule[1.5pt]
\end{tabular}
}{%
	\caption{Class-wise results for 4-class classification}
	\label{results4}
} 
\end{floatrow}
\end{figure}

\begin{figure}[!ht]
	\centering
	\subfloat[3-class classification]{\includegraphics[width=0.49\linewidth]{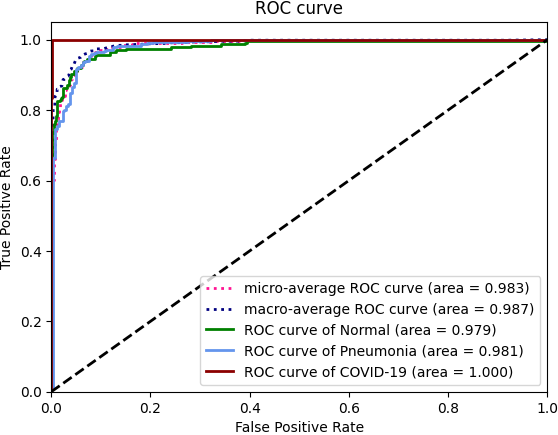}}
	\hfill
	\subfloat[4-class classification]{\includegraphics[width=0.49\linewidth]{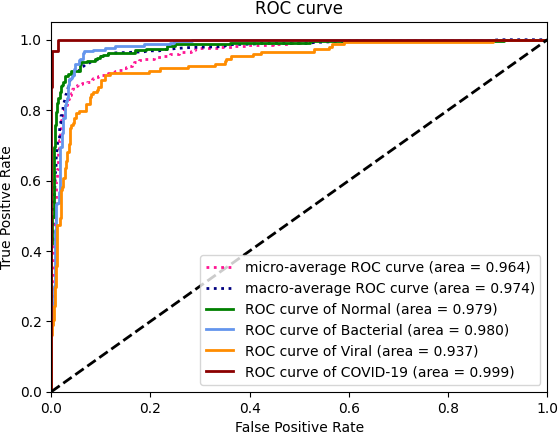}}

	\caption{ROC curves for our two configurations}
	\label{fig:roc_curves}
\end{figure}

We obtained an accuracy of $87.2\%$ for 4-class classification configuration and $90.5\%$ for the 3-class classification. The class-wise results for AUROC, Sensitivity (Recall) and PPV (Positive Predictive Value or Precision) are given in Tables \ref{results3} and \ref{results4}. The corresponding ROC curves are shown in Fig. \ref{fig:roc_curves}.

\begin{figure}[!ht]
	\centering
	\subfloat[3-class classification]{\includegraphics[width=0.45\linewidth]{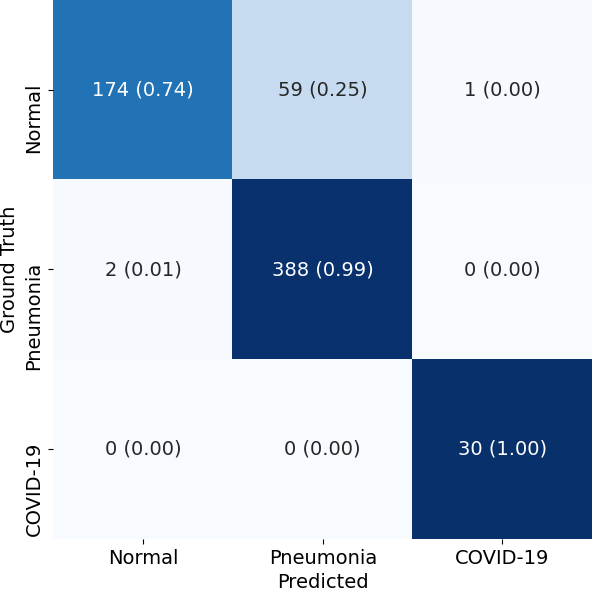}}
	\hfill
	\subfloat[4-class classification]{\includegraphics[width=0.45\linewidth]{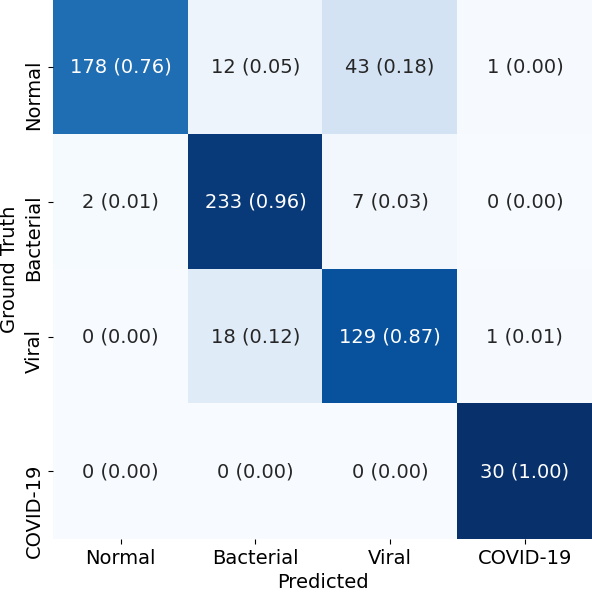}}

	\caption{Confusion matrices for our two configurations}
	\label{fig:conf_matrices}
\end{figure}

The confusion matrices on the test data for the two configurations of our model are shown in Fig. \ref{fig:conf_matrices}. It can be seen that viral pneumonia, having a sensitivity (recall) value of $0.87$ is often confused with bacterial pneumonia. This is likely due to the overlapping imaging characteristics. Our sensitivity (recall) for \covid positive class is $1.0$, which is at par with the sensitivity (recall) for bacterial pneumonia. 

\subsection{Comparison with COVID-Net}

\begin{figure}[!ht]
	\centering
	\subfloat[COVID-Net confusion matrix]{\includegraphics[width=0.45\linewidth, height=0.45\linewidth]{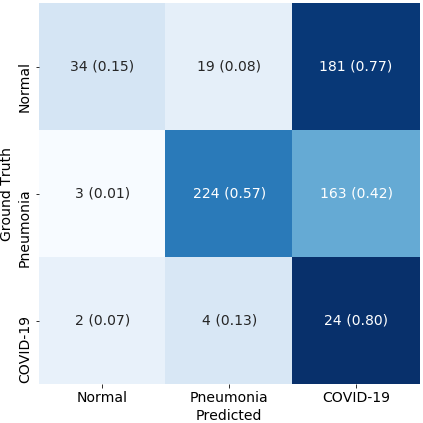}}
	\hfill
	\subfloat[ROC Curves: COVID-Net v/s Ours]{\includegraphics[width=0.54\linewidth]{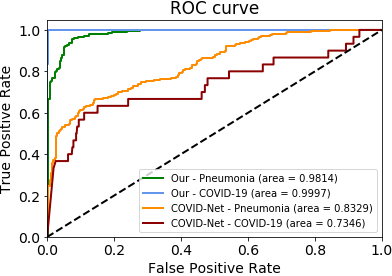}}

	\caption{Comparison with COVID-Net}
	\label{fig:covid-net_comparison}
\end{figure}

To compare our approach with COVID-Net~\cite{wang2020covidnet} we evaluate their published pre-trained models on the same test data split as ours. Fig. \ref{fig:covid-net_comparison}  compares their ROC curves with our model along with the confusion matrix. The COVID-Net inference shown here is made using their 'Small' variant which seemed to perform better among their two variants. It can be seen that our model outperforms COVID-Net by $>$0.14 AUROC in detecting regular Pneumonia as well as COVID-19. It should be noted that we used a different Pneumonia dataset from that used by COVID-Net, however, the COVID-19 data used is the same.

\begin{table}
	\setlength\tabcolsep{10pt}
	\centering
	\begin{tabular}{l c c}
	\toprule[1.5pt]
	Model & F1 Score & 95\% Confidence Interval\\
	\midrule[1pt]
	COVID-Net & 0.3591 & 0.3508 - 0.3674 \\
	CovidAID & 0.9230 & 0.9178 - 0.9282 \\
	\bottomrule[1.5pt]
\end{tabular}
	\caption{F1 Score (95\% CI) comparison}
	\label{f1_scores}
\end{table}

We compute F1 scores for CovidAID and Covid-Net~\cite{wang2020covidnet} over 10,000 bootstrapped samples from our test set of 654 images. 100 instances are taken (with  replacement) of size 100 each for the bootstrap. The F1 scores along with their 95\% confidence intervals are shown in Table \ref{f1_scores}.

The high margin of difference of the F1 scores of the two models clearly establish the superior performance of our model over COVID\hyp{}Net.

\subsection{Visualizations}

To demonstrate the results qualitatively, we generate saliency maps for our model's predictions using RISE~\cite{Petsiuk2018rise}. In this approach 1000 randomly masked versions of a given X-ray image are queried and their classification scores are used to create a weighted mask corresponding to each output class. The core idea behind the RISE~\cite{Petsiuk2018rise} approach is that masks which preserve semantically important parts of the image will lead to a higher classification score and hence a higher weight in the final mask for the respective class. 

\begin{figure}[h]
	\centering
	\captionsetup{justification=centering,margin=0.8cm}
	\subfloat{\includegraphics[width=0.3\linewidth]{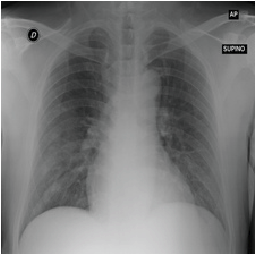}}
	\hfill
	\subfloat{\includegraphics[width=0.3\linewidth]{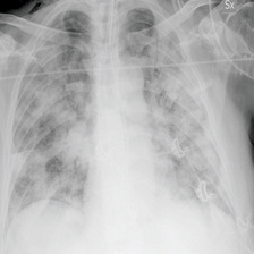}}
    \hfill
	\subfloat{\includegraphics[width=0.3\linewidth]{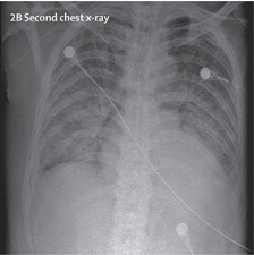}}
	\hfill
	\subfloat{\includegraphics[width=0.3\linewidth]{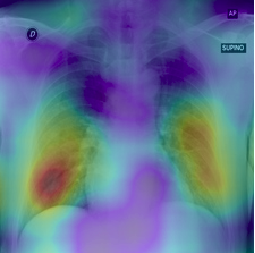}}
	\hfill
	\subfloat{\includegraphics[width=0.3\linewidth]{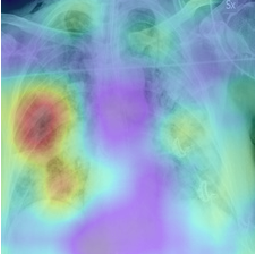}}
    \hfill
	\subfloat{\includegraphics[width=0.3\linewidth]{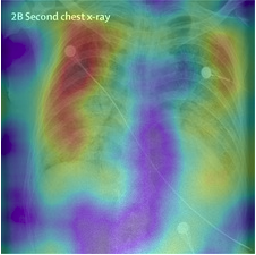}}
	\caption{Saliency map visaulization over COVID-19 positive X-Rays. Red denotes region of greater importance.}
	\label{fig:vis_images}
\end{figure}

The purpose of these visualizations was to have an additional check to rule out model over-fitting as well as to validate whether the regions of attention correspond to the right features from a radiologist's perspective.  Fig. \ref{fig:vis_images} shows some of the saliency maps on COVID-19 positive X-rays. 

\section{Conclusion}
We have presented some initial results on detecting COVID-19 positive cases from chest X-Rays using a deep-learning model. We have demonstrated significant improvement in performance over COVID-Net~\cite{wang2020covidnet}, the only publicly maintained tool for classification of COVID-19 positive X-rays, on the same chest-xray-pneumonia dataset \cite{kpneumonia}. The results look promising, though the size of the publicly available dataset is small. We plan to further validate our approach using larger  \covid X-ray image datasets and clinical trials. 

\bibliographystyle{splncs04}
\bibliography{ref}

\end{document}